\documentclass[aps,pra,twocolumn,amsmath,amssymb,nofootinbib,superscriptaddress]{revtex4-1}

\newcommand{\bra}[1]{\left\langle#1\right\rvert}
\newcommand{\ket}[1]{\left\lvert#1\right\rangle}
\newcommand{\ketbra}[2]{\left\lvert{#1}\middle\rangle\!\middle\langle{#2}\right\rvert}
\newcommand{\braket}[2]{\left\langle{#1}\middle\vert{#2}\right\rangle}
\newcommand{\expect}[1]{\left\langle{#1}\right\rangle}
\newcommand{\tr}{\mathrm{Tr}}
\newcommand{\set}[1]{\left\lbrace #1 \right\rbrace}

\usepackage[colorlinks,linkcolor=blue,citecolor=blue,allcolors=blue]{hyperref} 
\usepackage[pdftex]{graphicx}
\usepackage{subfigure}
\usepackage{mathrsfs}
\usepackage{subfigure}
\usepackage{float}
\usepackage[usenames,dvipsnames]{xcolor}
\usepackage{framed}

\begin{document}

\title{Optimal Fidelity Witnesses for Gravitational Entanglement}

\author{Thomas Guff}
\email[]{thomas.guff@fysik.su.se}
\affiliation{Department of Physics, Stockholm University, SE-106 91 Stockholm, Sweden}

\author{Nicolas Boulle}
\email[]{nicolasboulle11@gmail.com}
\affiliation{Department of Physics, Stockholm University, SE-106 91 Stockholm, Sweden}

\author{Igor Pikovski}
\email[]{igor.pikovski@fysik.su.se}
\affiliation{Department of Physics, Stockholm University, SE-106 91 Stockholm, Sweden}
\affiliation{Department of Physics, Stevens Institute of Technology, Hoboken, New Jersey 07030, USA}

\date{\today}

\begin{abstract}
    Optomechanical systems open new possibilities in fundamental research at the interface between quantum information and gravity. 
    Recently, an ambitious experimental proposal was suggested by Bose \emph{et al.} \cite{Bose2017} to measure the entanglement between two optomechanical systems generated by their gravitational interaction. 
    The scheme relied on witnessing entanglement between the two systems. Here we develop a general framework to study the quality of bipartite entanglement witnesses using fidelity witnesses. 
    We then apply this framework to the gravitational entanglement proposal, optimizing for the detection of entanglement. 
    We construct a witness consisting of only five non-trivial spin measurements, which we compare with other proposed witnesses.
    With post-processing our witness can detect entanglement for any choice of phases in the setup, up to a set of measure zero, for a closed system. We also explore the effects of a simple dephasing channel on this witness.
\end{abstract}

\maketitle

\section{Introduction}\label{sec:intro}

Quantum entanglement is considered one of the principal features of quantum mechanics that sets it apart from classical physics.
It is the crucial ingredient for quantum systems to violate Bell's inequalities, demonstrating that there can be no local hidden variable theory to reproduce quantum measurement statistics \cite{Bell1964,Brunner2014,Horodecki2009}. Pushing the study of quantum effects into the mesoscopic realm, entanglement is now studied also in the context of optomechanical systems, which promise the optical quantum control and read-out of massive mechanical cantilevers or levitated spheres \cite{aspelmeyer2014cavity}. Optomechanics can  generate entanglement between light and a massive oscillator \cite{ghobadi2014optomechanical,marinkovic2018optomechanical}, as well as between two distant massive resonators \cite{joshi2012entanglement,clarke2020generating}. Optomechanics also opens new possibilities to study gravity in the quantum regime \cite{marshall2003towards, kleckner2008creating, pikovski2012probing, belenchia2016testing}. One class of experimental proposals relies on entanglement generation between two gravitating systems that can shed light on the quantum character of gravity  \cite{Bose2017,Marletto2017,Carney2021,Pedernales2021,Ma2022,Hosten2022,streltsov2022,Cosco2021}. In particular, the proposal by Bose \emph{et al.} \cite{Bose2017} suggested creating superpositions of two levitated spheres near each other, such that their gravitational interactions caused entanglement between them that can be read out using an internal spin system. While there has been impressive experimental progress to study the gravitational interaction between classical oscillators \cite{westphal2021measurement}, realizing entanglement is experimentally very challenging given the weak nature of the gravitational force between objects of very small mass, and the difficulty in maintaining massive objects in a spatial superposition for extended durations \cite{rijavec2021decoherence, tilly2021qudits}. It is thus necessary to optimize possible measurement protocols for verifying entanglement between the two masses \cite{Chevalier2020}.

Entanglement witnesses are helpful tools to experimentally verify  entanglement. An entanglement witness is an observable which can demonstrate that a given system is entangled, if the expectation value of that observable lies within a particular range of values \cite{Horodecki2009,Horodecki1996}.
For any particular entangled state, one can always find an observable to serve as an entanglement witness \cite{Horodecki1996}. However, not all entanglement witnesses are created equal. Many will only detect entanglement for a small number of entangled states.
Furthermore, in real experiments, measurements of entanglement witnesses will be done by performing local measurements on quantum systems; for example, local spin measurements. Thus there is a preference for observables which do not require as many measurements as would be required to do full-state tomography \cite{Cramer2010,Resch2005}.

A particular class of entanglement witnesses are fidelity witnesses \cite{Bourennane2004,Gluza2018}. They are constructed from a particular entangled state and positive multiple of the identity operator. This proportionality constant provides a natural measure for the volume of entangled states for which the fidelity witness will detect entanglement.
Unsurprisingly, fidelity witnesses constructed from maximally entangled states yield the widest detection areas.
Most research on fidelity witnesses has been focused on trying to find optimal entanglement measures \cite{Lewenstein2000,Lewenstein2001,Eisert2004,Hyllus2006} and in particular, the smallest representation of a given witness in terms of local measurements \cite{Bourennane2004,Riccardi2020,Guhne2002,Guhne2003}. 

In this paper, we use the framework of fidelity witnesses to measure the quality of bipartite entanglement witnesses. By generalising the framework to witnesses constructed out of mixed states, we are able to measure the detection volume of a much broader set of entanglement witnesses.
From this we argue that the optimal entanglement witness is one constructed from a closest maximally entangled state to the state of interest. 

We apply this framework to the experimental proposal of Bose \emph{et al.} \cite{Bose2017} to develop an entanglement witness containing just five non-trivial local spin measurements; compared with the 15 required for full state tomography. With the appropriate post-processing this witness can detect entanglement for any given set of phases in the given experimental proposal, except for a set of measure zero. It can thus cover the entire configuration space and can be optimised in post-processing for a given system configuration.

This paper is structured as follows. In Sec.~\ref{sec:ew} we provide a brief overview of entanglement witnesses, in particular fidelity witnesses, and generalise the construction to include a wider range of witnesses. In Sec.~\ref{sec:optimal}, we argue for the optimal fidelity witness relative to a given pure state, and show how to construct it. In Sec.~\ref{sec:gw} we apply it to a recent experimental proposal \cite{Bose2017} to detect gravitational induced entanglement, leading to a an improved entanglement witness. In Sec.~\ref{sec:dephasing} we look at how well this witness detects entanglement in the presence of a simple dephasing channel. Finally in Sec.~\ref{sec:disc} we discuss the generality of our constructions and how they could be generalised further.

\section{Entanglement Witnesses}\label{sec:ew}

An entanglement witness is an observable used to confirm that the state of a physical system is entangled. In particular, an observable $W$ is an entanglement witness if there exists a set $\mathcal{X}\subseteq \mathbb{R}$ of real numbers such that for any quantum state $\ket{\psi}$, if $\bra{\psi}W\ket{\psi} \in \mathcal{X}$ then $\ket{\psi}$ is entangled.
We call this set $\mathcal{X}$ the \emph{set of detection values}.

In this paper we will consider entanglement on bipartite systems which are finite dimensional. 
We also focus attention on entanglement witnesses for which $\mathcal{X} = \set{x\in \mathbb{R} \,|\, x<0 }$.
In other words we consider observables $W$ on bipartite systems $\mathcal{H}_{A}\otimes \mathcal{H}_{B}$ for which $\bra{\psi}\bra{\phi}W\ket{\psi}\ket{\phi} \geq 0$ for all product states $\ket{\psi}\ket{\phi}$, but is not positive semi-definite, i.e. $W$ has at least one negative eigenvalue.
This has the advantage that the set of states for which $W$ will register entanglement is unchanged if we rescale $W$ by a positive real number. 
In Sec.~\ref{sec:disc} we discuss how to generalise to witnesses which do not meet this criteria.

\subsection{Fidelity Witnesses}\label{sec:fw}

Let $\mathcal{H}_{A}\otimes \mathcal{H}_{B}$ be a finite-dimensional Hilbert space. 
Suppose we have a bipartite entangled pure state $\ket{\psi} \in \mathcal{H}_{A}\otimes \mathcal{H}_{B}$.
We can construct an entanglement witness specifically using this state (and that is useful for nearby states). Explicitly, this is given by
\begin{equation}
W_{\psi}= \alpha \mathbb{I} - \ketbra{\psi}{\psi},\label{eq:W}
\end{equation}
where $\mathbb{I}$ is the identity operator, and $\alpha$ is defined by
\begin{equation}
\alpha = \max_{\ket{\phi}\in \mathcal{S}} |\braket{\phi}{\psi}|^{2} \label{eq:alpha}
\end{equation}
where $\mathcal{S}$ is the set of all product states of $\mathcal{H}_{A}\otimes \mathcal{H}_{B}$.
The coefficient $\alpha$ is equal to $1$ if and only if $\ket{\psi}$ is a product state. If $\ket{\psi}$ is entangled then  $\alpha < 1$.

The observable $W_{\psi}$ constructed in the manner of $\eqref{eq:W}$ is known as a \emph{fidelity witness} \cite{Bourennane2004,Guhne2009}. The observable $W_{\psi}$ is clearly an entanglement witness since if $\ket{\phi}\in \mathcal{S}$, then
\begin{equation}
    \bra{\phi}W_{\psi}\ket{\phi} = \alpha - |\braket{\phi}{\psi}|^{2} \geq 0.
\end{equation}
but it is not positive-definite, since
\begin{equation}
    \bra{\psi}W_{\psi}\ket{\psi} = \alpha - 1 < 0.
\end{equation}
Furthermore, if $\rho$ is a separable state, then it can be written as
\begin{equation}
    \rho = \sum_{i=1}^{n} p_{i} \ketbra{a_{i}}{a_{i}} \otimes \ketbra{b_{i}}{b_{i}}.
\end{equation}
For any separable state $\rho$ we have
\begin{equation}
    \tr(W_{\psi}\rho) = \sum_{i=1}^{n} p_{i} \bra{a_{i}}\bra{b_{i}}W_{\psi}\ket{a_{i}}\ket{b_{i}} \geq 0,
\end{equation}
therefore, if $\tr(W_{\psi}\rho)<0$ for any mixed state $\rho$ then $\rho$ is guaranteed to be entangled.

The number $1-\alpha$ can be thought of as measuring the ``distance" of $\ket{\psi}$ to the nearest product state. 
It can also be seen as measuring the size of the region around $\ket{\psi}$ in the space of states for which $W$ will confirm entanglement.
This is because $\tr(W\rho)<0$ if and only if $\bra{\psi}\rho\ket{\psi} > \alpha$, so the witness $W$ will detect entanglement for all and only those states $\rho$ whose fidelity with $\ket{\psi}$ is greater than $\alpha$.

We wish to calculate an explicit expression for $\alpha$. Suppose $d=\dim(\mathcal{H}_{A}) \leq \dim(\mathcal{H}_{B})=d^{\prime}$. Then we can write $\ket{\psi}$ using the Schmidt decomposition \cite{Nielsen2010},
\begin{equation}
    \ket{\psi} = \sum_{i=1}^{d} \sqrt{p_{i}} \ket{\varphi_{i}^{A}}\ket{\varphi_{i}^{B}}.
\end{equation}
where $\set{\ket{\varphi^{A}_{i}}}$ and $\set{\ket{\varphi^{B}_{i}}}$ are sets of orthonormal states and $p_{i}\geq 0$. As shown in \cite{Bourennane2004}, we can use this to calculate $\alpha$. Let $\ket{\phi}$ be a product state, and extend $\set{\ket{\varphi^{B}_{i}}}$ to be an orthonormal basis for $\mathcal{H}_{B}$. We can write
\begin{equation}
    \ket{\phi} = \sum_{i=1}^{d}\sum_{j=1}^{d^{\prime}} a_{i}b_{j} \ket{\varphi_{i}^{A}}\ket{\varphi_{j}^{B}}.
\end{equation}
We then have
\begin{align}
   |\braket{\phi}{\psi}|^{2} &= \left|\sum_{i=1}^{d} \sqrt{p_{i}} a_{i}^{*}b_{i}^{*}\right|^{2} \leq \left(\sum_{i=1}^{d} \sqrt{p_{i}} |a_{i}||b_{i}|\right)^{2} \nonumber \\
    &\leq \left(\max_{i=1,\dots,d} \sqrt{p_{i}}\right)^{2} = \max_{i=1,\dots,d} p_{i},
\end{align}
where the second inequality follows since $|a_{i}|,|b_{i}|\leq 1$.
So we see that $\alpha$ is bounded by the square of the largest Schmidt coefficient of $\ket{\psi}$. But this bound can be reached since, if $p_{n} = \max_{i} p_{i}$, then
\begin{equation}
   \left|\left(\bra{\varphi_{n}^{A}}\bra{\varphi_{n}^{B}}\right)\ket{\psi}\right|^{2} = p_{n},
\end{equation}
and therefore
\begin{equation}
    \alpha = \max_{i=1,\dots, d} p_{i}.
\end{equation}
Therefore, we can conclude from this that the smallest value $\alpha$ can take for any state is $\alpha = 1/\dim(\mathcal{H}_{A})$, which occurs when $\ket{\psi}$ is maximally entangled. So, for example, in the case of two qubits, $\alpha \geq 1/2$.
 
\subsection{Generalisation}\label{sec:generalise}

Clearly not all entanglement witnesses have the form of \eqref{eq:W}, since an operator of this form must have only one negative eigenvalue and all positive eigenvalues equal. However we can put any entanglement witness which detects entanglement with a negative expectation value into a form analogous to \eqref{eq:W}, but constructed from a mixed state. First note that any Hermitian operator $W$ which is not proportional to the identity, can be uniquely written in the form
\begin{equation}
    W = \beta \mathbb{I} - \left[\beta\delta-\tr(W)\right]rho,
\end{equation}
where $\rho$ is a positive semi-definite operator with unit trace, $\beta$ is the largest eigenvalue of $W$ and $\delta=\tr(\mathbb{I})=\dim(\mathcal{H}_{A}\otimes \mathcal{H}_{B})$. 
To see this, first note that if $W$ is not proportional to the identity, then $\beta \delta - \tr(W)>0$, and therefore
\begin{equation}
    \rho = \frac{\beta \mathbb{I} - W}{\beta \delta - \tr(W)},
\end{equation}
is uniquely defined and clearly has unit trace. 
That $\rho$ is positive semi-definite follows from the fact that for all $\ket{\phi} \in \mathcal{H}_{A}\otimes \mathcal{H}_{B}$, $\beta \geq \bra{\phi}W\ket{\phi}$ since $\beta$ is the largest eigenvalue of $W$.

If $W$ is an entanglement witness then it must have at least one negative and one non-negative eigenvalue, and therefore, it cannot be proportional to the identity and also $\beta \geq 0$.

Since an entanglement witness is functionally unchanged under rescaling by a positive constant (it does not change the set of states that have a negative expectation value), and since $\beta \delta > \tr(W)$, we can consider the operator
\begin{equation}
    W^{\prime} = \frac{W}{\beta \delta - \tr(W)} = \alpha\mathbb{I} - \rho, \label{eq:rescaled}
\end{equation}
where clearly $\alpha=\beta/\left[\beta \delta-\tr(W)\right]$. Since $W$ is an entanglement witness and has at least one negative eigenvalue, $\tr(W) < \beta(\delta-1)$, from which it follows that $\beta \delta - \tr(W) > \beta$, hence $0 \leq \alpha < 1$. Clearly $\alpha$ must be in this range otherwise \eqref{eq:rescaled} could not be an entanglement witness.

Note that since $W$ is general, $\alpha$ need not be the maximal overlap of $\rho$ with any product state, analogous to \eqref{eq:alpha}. Indeed in \eqref{eq:W}, $\alpha$ can be increased (as long as it does not reach unity) and $W$ will still remain an entanglement witness, albeit a weaker one. But $\alpha$ cannot be smaller than the maximal overlap with any product state
\begin{equation}
    \alpha \geq \max_{\ket{\phi} \in S} \bra{\phi}\rho\ket{\phi}.
\end{equation}

Thus all entanglement witnesses (which register entanglement with a negative expectation value) can be written in an analogous form to \eqref{eq:W} (after a suitable rescaling).
Note, for such an entanglement witness, we have
\begin{equation}
    \bra{\phi}W^{\prime}\ket{\phi} = \alpha - \bra{\phi}\rho\ket{\phi} \geq \alpha - \lambda,
\end{equation}
where $\lambda$ is the largest eigenvalue of $\rho$. 
Thus we can use $\lambda - \alpha$ as a measure of the region of states around $\rho$ for which $W^{\prime}$ will detect entanglement.

From \eqref{eq:rescaled} we see that for a general density matrix $\sigma$, $W^{\prime}$ will register entanglement if and only if $\tr(\rho\sigma) > \alpha$.

\section{Optimal Fidelity Witness}\label{sec:optimal}

The purpose of an entanglement witness is to confirm the entanglement of a particular state of interest $\ket{\psi}$. The task then is to choose the optimal fidelity witness to measure such entanglement. It might seem obvious to choose the fidelity witness as in \eqref{eq:W}, constructed from the state of interest $\ket{\psi}$. However if the state $\ket{\psi}$ has very little entanglement, then $\alpha$ will be close to unity, and the region of detection will have small size $1-\alpha$. This means that small experimental errors causing the state to deviate slightly from $\ket{\psi}$ may result in $W$ not detecting the entanglement. 
Indeed the exact target state might not be kown \emph{a priori}.
Thus we value a large detection area.

We saw in Sec.~\ref{sec:generalise}, that the size of the region for which a fidelity witness will detect entanglement was best measured by $\lambda - \alpha$. Thus to maximise this area, we wish to maximise $\lambda$ and minimise $\alpha$. This can be done simultaneously by choosing a fidelity witness of the form $\eqref{eq:W}$, where $\ket{\psi}$ is a maximally entangled state.

Given an entangled state $\ket{\phi}$, we define an \emph{optimal fidelity witness for $\ket{\phi}$} as the fidelity witness constructed from a maximally entangled state closest to $\ket{\phi}$, as measured by the fidelity.

Again assume $d=\dim(\mathcal{H}_{A}) \leq \dim(\mathcal{H}_{B})=d^{\prime}$. A state is defined to be maximally entangled, if it has a Schmidt number (the number of non-zero terms in the Schmidt decomposition) of $d$, and if all Schmidt coefficients are equal.

To find the closest maximally entangled state to a particular state $\ket{\psi}$, we first write $\ket{\psi}$ in its Schmidt decomposition
\begin{equation}
\ket{\psi} = \sum_{i=1}^{d} \sqrt{p_{i}}\ket{\psi_{i}^{A}}\ket{\psi_{i}^{B}} \label{eq:schmidt}
\end{equation}
where $\set{\ket{\psi^{A}_{i}}}_{i=1}^{d}$ and $\set{\ket{\psi^{B}_{i}}}_{i=1}^{d^\prime}$ are orthonormal bases.
We now search for a maximally entangled state with the largest inner product with $\ket{\psi}$.
Let us consider a general state
\begin{equation}
\ket{\theta} = \sum_{i=1}^{d}\sum_{j=1}^{d^{\prime}} a_{ij}\ket{\psi^{A}_{i}}\ket{\psi^{B}_{j}}.
\end{equation}
Then
\begin{align}
|\braket{\psi}{\theta}|^{2} &=\left|\sum_{i,k=1}^{d}\sum_{j=1}^{d^{\prime}}\sqrt{p_{k}} a_{ij}\braket{\psi_{k}^{A}}{\psi^{A}_{i}}\braket{\psi^{B}_{k}}{\psi^{B}_{j}}\right|^{2} \nonumber \\
&= \left| \sum_{i,k=1}^{d}\sum_{j=1}^{d^{\prime}}\sqrt{p_{k}} a_{ij}\delta_{k,i}\delta_{k,j}\right|^{2}. \label{eq:ip}
\end{align}
Thus we see that the `cross' terms $a_{ij}$ where $i\ne j$ are actively reducing the inner product, since they contribute nothing to the inner product themselves but they reduce the size of the `diagonal' terms. So to maximise \eqref{eq:ip} we should set $a_{ij}=0$ if $i\ne j$. After this we are compelled to choose $a_{ii} = e^{i\varphi_{i}}/\sqrt{d}$, so that $\ket{\theta}$ is maximally entangled. So we have
\begin{equation}
    |\braket{\psi}{\theta}|^{2} = \left| \sum_{i=1}^{d} \sqrt{p_{i}} \frac{e^{i\varphi_{i}}}{\sqrt{d}}\right|^{2} = \frac{1}{d}\sum_{i,j=1}^{d} 
    \sqrt{p_{i}p_{j}}\cos(\varphi_{i}-\varphi_{j}).
\end{equation}
This sum is maximised when $\cos(\varphi_{i}-\varphi_{j})=1$ for all $\sqrt{p_{i}p_{j}}\ne 0$. This implies that up to a global phase
\begin{equation}
    e^{i\varphi_{i}} = 1 \quad \text{if} \quad p_{i} \ne 0. \label{eq:phasecond}
\end{equation}
So we have a phase freedom for all terms in the Schmidt decomposition corresponding to terms where $p_{i}=0$. Thus the closest maximally entangled state to $\ket{\psi}$ is 
\begin{equation}
\ket{\theta} = \sum_{i=1}^{d} \frac{e^{i\varphi_{i}}}{\sqrt{d}}\ket{\psi_{i}^{A}}\ket{\psi_{i}^{B}}. \label{eq:closest}
\end{equation}
where the phase terms must satisfy \eqref{eq:phasecond}.
We see that the closest maximally entangled state is uniquely determined if the Schmidt number of $\ket{\psi}$ (the number of non-zero terms in the Schmidt decomposition) is $d$, and not otherwise. In particular the closest maximally entangled state to a product state is not unique.

If we now construct the fidelity witness corresponding to the state $\ket{\theta}$ we find,
\begin{equation}
W_{\theta} = \frac{1}{d}\mathbb{I} - \ketbra{\theta}{\theta}.\label{eq:thetaGens}
\end{equation}
If we calculate the expectation value with respect to the original state of interest $\ket{\psi}$, we find
\begin{align}
\expect{W_{\theta}}_{\psi} = \frac{1}{d}-|\braket{\psi}{\theta}|^{2}&=\frac{1}{d}-\left| \sum_{i=1}^{d} \frac{\sqrt{p_{i}}}{\sqrt{d}}\right|^{2} \nonumber \\
&= -\frac{1}{d}\sum_{i\ne j}^{d} \sqrt{p_{i}p_{j}}.
\end{align}

This derivation required initially decomposing the state of interest into a Schmidt decomposition \eqref{eq:schmidt}, so we must discuss the uniqueness of the Schmidt decomposition.
The Schmidt decomposition is generally not unique, although the Schmidt coefficients $\sqrt{p_{i}}$ are unique. If $\set{\ket{\psi^{A}_{i}}\ket{\psi^{B}_{i}}}_{i=1}^{n}$ are terms in the Schmidt decomposition which appear with the same Schmidt coefficient, then they can be replaced by $\set{\ket{\phi^{A}_{i}}\ket{\phi^{B}_{i}}}_{i=1}^{n}$, where $\set{\ket{\phi^{A}_{i}}}$ and $\set{\ket{\phi^{B}_{i}}}$ are orthonormal sets, if
\begin{equation}
   \sum_{i=1}^{n} \ket{\psi^{A}_{i}}\ket{\psi^{B}_{i}} = \sum_{i=1}^{n} \ket{\phi^{A}_{i}}\ket{\phi^{B}_{i}}.
\end{equation}
Such a replacement will leave the closest maximally entangled state \eqref{eq:closest} unchanged.
If the Schmidt decomposition has any terms corresponding to a zero coefficient, then there is an additional phase freedom, as $\ket{\psi^{A}_{i}}\ket{\psi^{B}_{i}}$ can be replaced with $\ket{\phi^{A}_{i}}\ket{\phi^{B}_{i}} = e^{i\varphi}\ket{\psi^{A}_{i}}\ket{\psi^{B}_{i}}$. However, this freedom is absorbed by the phase freedom in the closest maximally entangled state  \eqref{eq:phasecond}.

\section{Witnesses for Gravitational Entanglement}\label{sec:gw}

\begin{figure}[t]
    \centering
    \includegraphics[scale=0.4]{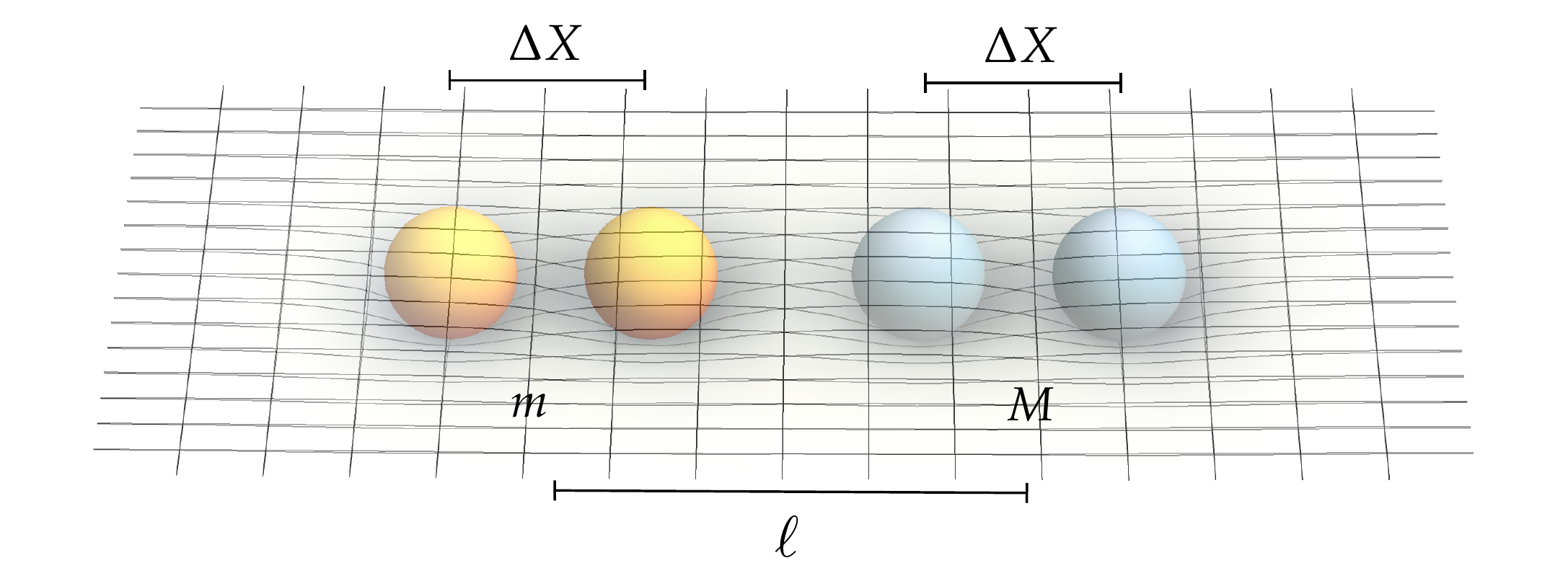}
    \caption{Diagrammatic setup of the experimental proposal by Bose \emph{et al.} \cite{Bose2017} described in Sec.~\ref{sec:gw}. Two masses, labelled $m$ and $M$ are both placed in superposition of sizes $\Delta X$ with a distance between their respective centres of mass $\ell$. The mutual Newtonian interaction induces entanglement between the two masses.}
    \label{fig:bose}
\end{figure}
The experiment proposed in \cite{Bose2017} consists of two masses, which we will label $m$ and $M$, both placed nearby in a superposition of length $\Delta X$, with a distance $\ell$ between their respective centres of mass (see Fig.~\ref{fig:bose} for a diagram). The masses are fixed in place, but the Newtonian interaction for a time $\tau$ will induce a relative phase between the component states.

After the gravitational interaction, the state is given by
\begin{align}
\ket{\psi(\phi_{1},\phi_{2})} &= \frac{1}{2} \left(\ket{m_{L}}\ket{M_{L}} + e^{i\phi_{1}}\ket{m_{L}}\ket{M_{R}}\right. \nonumber \\
&\quad\quad\left.+\,  e^{i\phi_{2}}\ket{m_{R}}\ket{M_{L}} + \ket{m_{R}}\ket{M_{R}} \right),  \label{eq:psi}
\end{align}
where the subscripts $L$ and $R$ denote the mass being in the ``left" or ``right" branch of the respective superposition.
In this particular experiment, the phases are determined by
\begin{subequations}\label{eq:phases}
\begin{align}
    \phi_{1} &= \frac{G M m\tau }{\hbar}\left(\frac{1}{\ell+\Delta X} - \frac{1}{\ell}\right), \\
    \phi_{2} &=  \frac{G M m\tau }{\hbar}\left(\frac{1}{\ell-\Delta X} - \frac{1}{\ell}\right),
\end{align}
\end{subequations}
however we will leave them general in this article.
The state $\ket{\psi(\phi_{1},\phi_{2})}$ is a product state when $\phi_{1} + \phi_{2} = 2 n\pi$ where $n\in \mathbb{Z}$ and not otherwise.
Furthermore, it is maximally entangled if and only if $\phi_{1}+\phi_{2} = (2n+1)\pi$.

In \cite{Bose2017}, the authors suggested the entanglement witness
\begin{equation}
    W_{1} = X\otimes Z + Y\otimes Y,
\end{equation}
which registers entanglement whenever ${|\expect{W}|>1}$. In other words the set of detection values is $\mathcal{X}=\set{x\in \mathbb{R} \, | \, x<-1 \; \text{and}\; x>1}$.
This witness is therefore not of the form we require in Sec.~\ref{sec:ew}. However, as in \cite{Chevalier2020}, we can ignore half of the set of detection values and put it into the form we required by adding the identity,
\begin{equation}
    W_{1}^{\prime} = I\otimes I + X\otimes Z + Y\otimes Y.\label{eq:W1p}
\end{equation}
Here we use $I$ to denote the identity operator on a subsystem, while retaining $\mathbb{I}$ to refer to the identity on the joint system.
We can justify this modification \eqref{eq:W1p} as this particular half of the set of detection values is the relevant one for the experimental values given in \cite{Bose2017}.
The observable $W_{1}^{\prime}$ has a largest eigenvalue of $\beta=3$, and therefore $\beta d - \tr(W_{1}^{\prime}) = 8$. We now write this witness in the form of \eqref{eq:rescaled}
\begin{equation}
    \frac{1}{8} W^{\prime}_{1} = \frac{3}{8}\mathbb{I} - \rho,
\end{equation}
where
\begin{equation}
    \rho = \frac{1}{4}I\otimes I - \frac{1}{8} X\otimes Z - \frac{1}{8} Y\otimes Y.
\end{equation}
The maximal eigenvalue of $\rho$ is $\frac{1}{2}$, therefore we can assign the witness $W^{\prime}_{1}$ a detection area of $\frac{1}{2}-\frac{3}{8}=\frac{1}{8}$ as described in Sec.~\ref{sec:generalise}. Thus we see that this witness has a sub-optimal detection area, compared to the maximal possible area of $1/2$. 
Furthermore, this witness is not useful in detecting small entanglement. Indeed, as pointed out in \cite{Chevalier2020}, for the experimental parameters given in \cite{Bose2017}, this witness does not detect entanglement until $\tau=8$s. Superpositions of this size and duration are extremely difficult to create, and are outside the envisioned parameter range of the experiment.
Thus a more suitable witness would be one that is sensitive to very small entanglement.

\begin{figure*}[t!]
    \centering
        \subfigure[]{\includegraphics[scale=0.32]{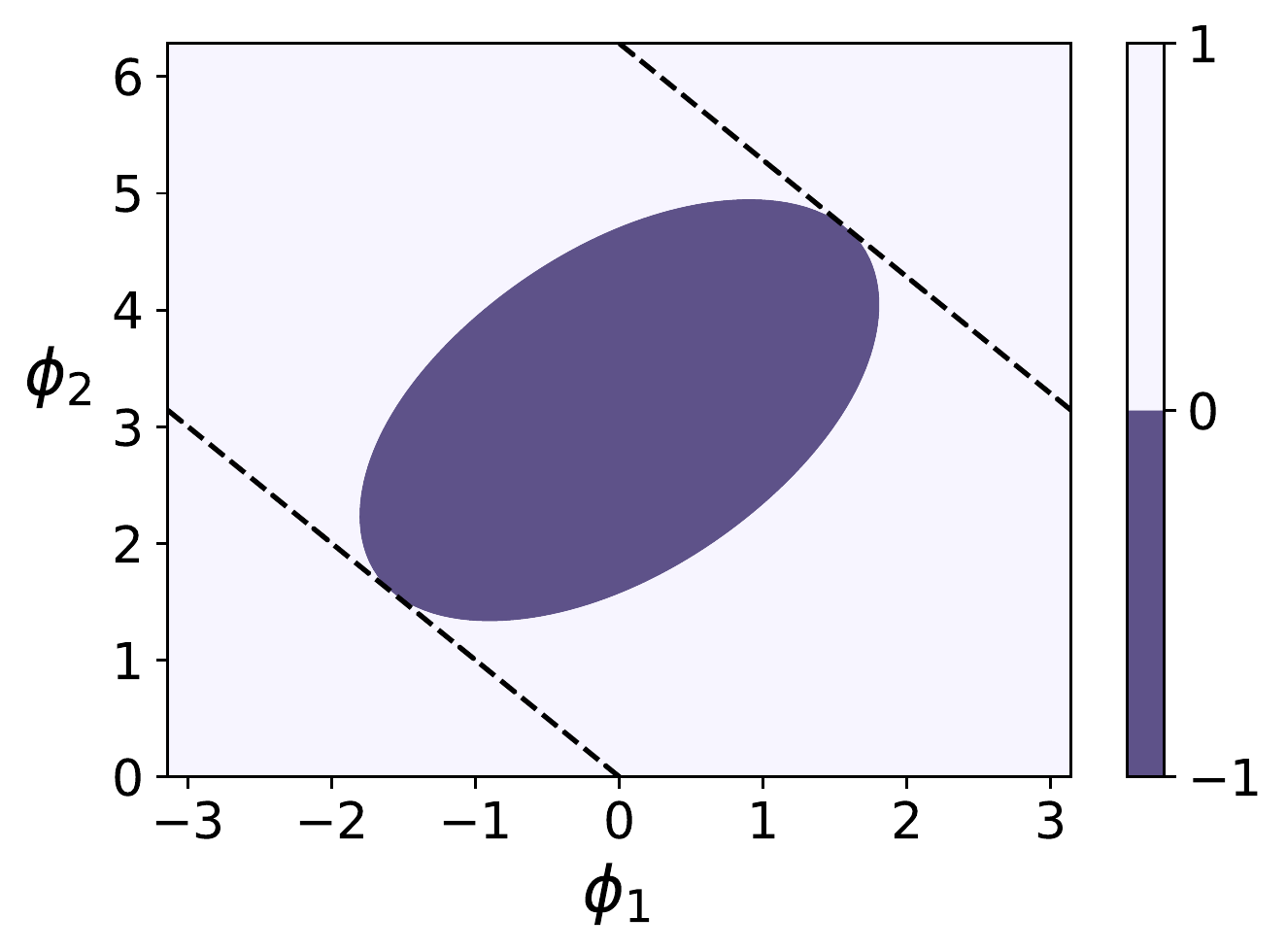}\label{fig:WBose}}
        \subfigure[]{\includegraphics[scale=0.32]{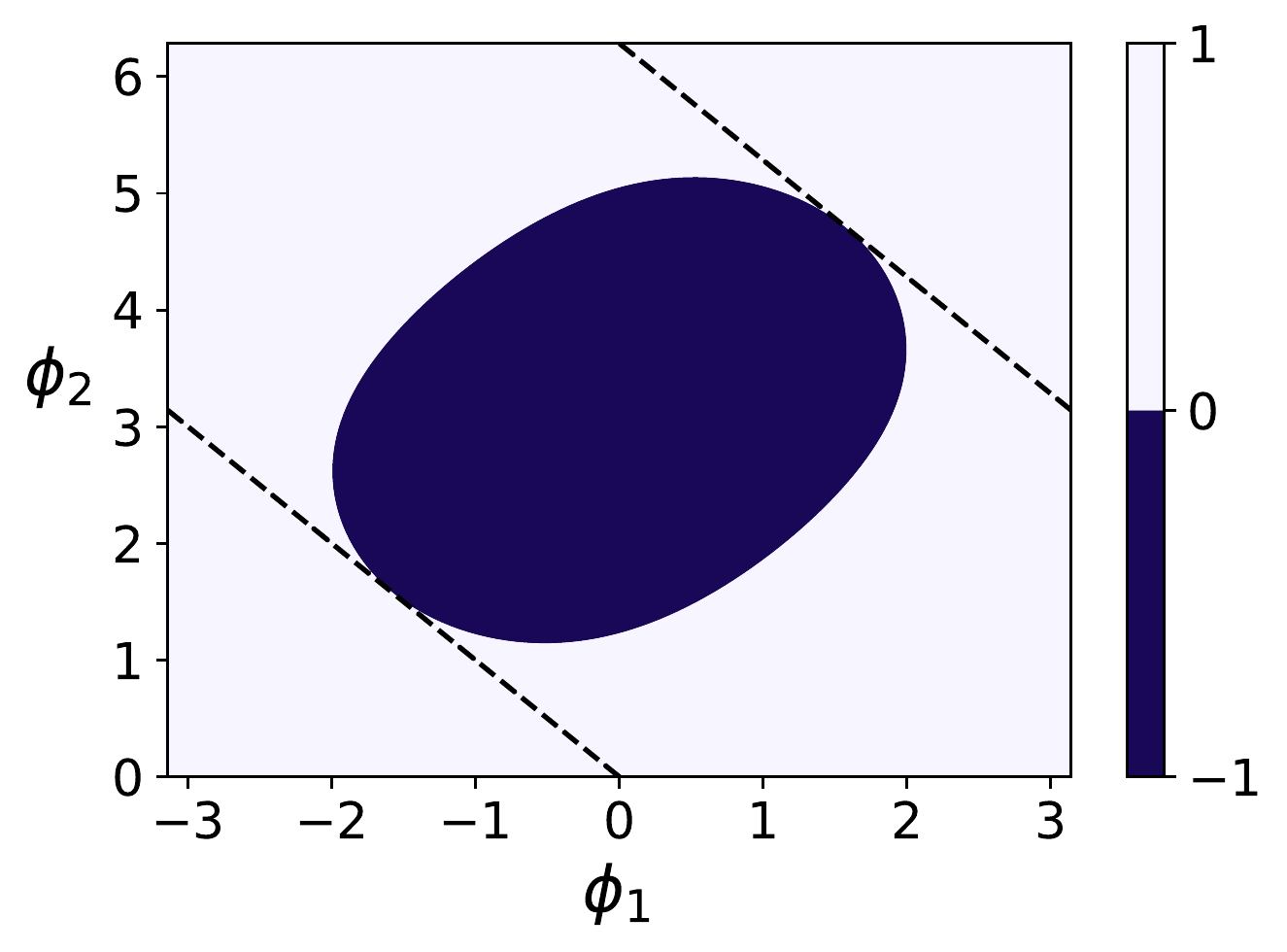}\label{fig:Wprime}}
        \subfigure[]{\includegraphics[scale=0.32]{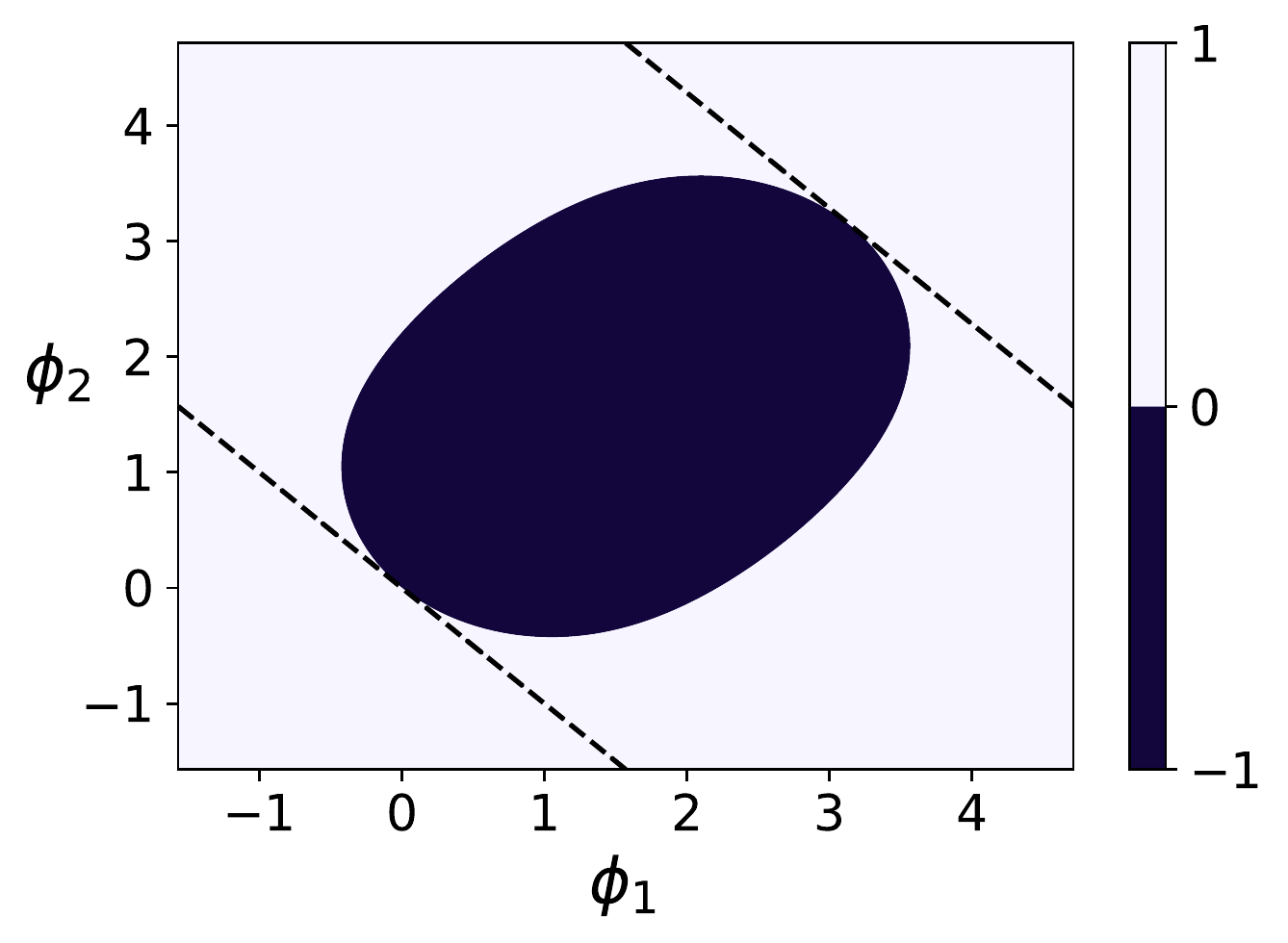}\label{fig:WChev}}
        \subfigure[]{\includegraphics[scale=0.32]{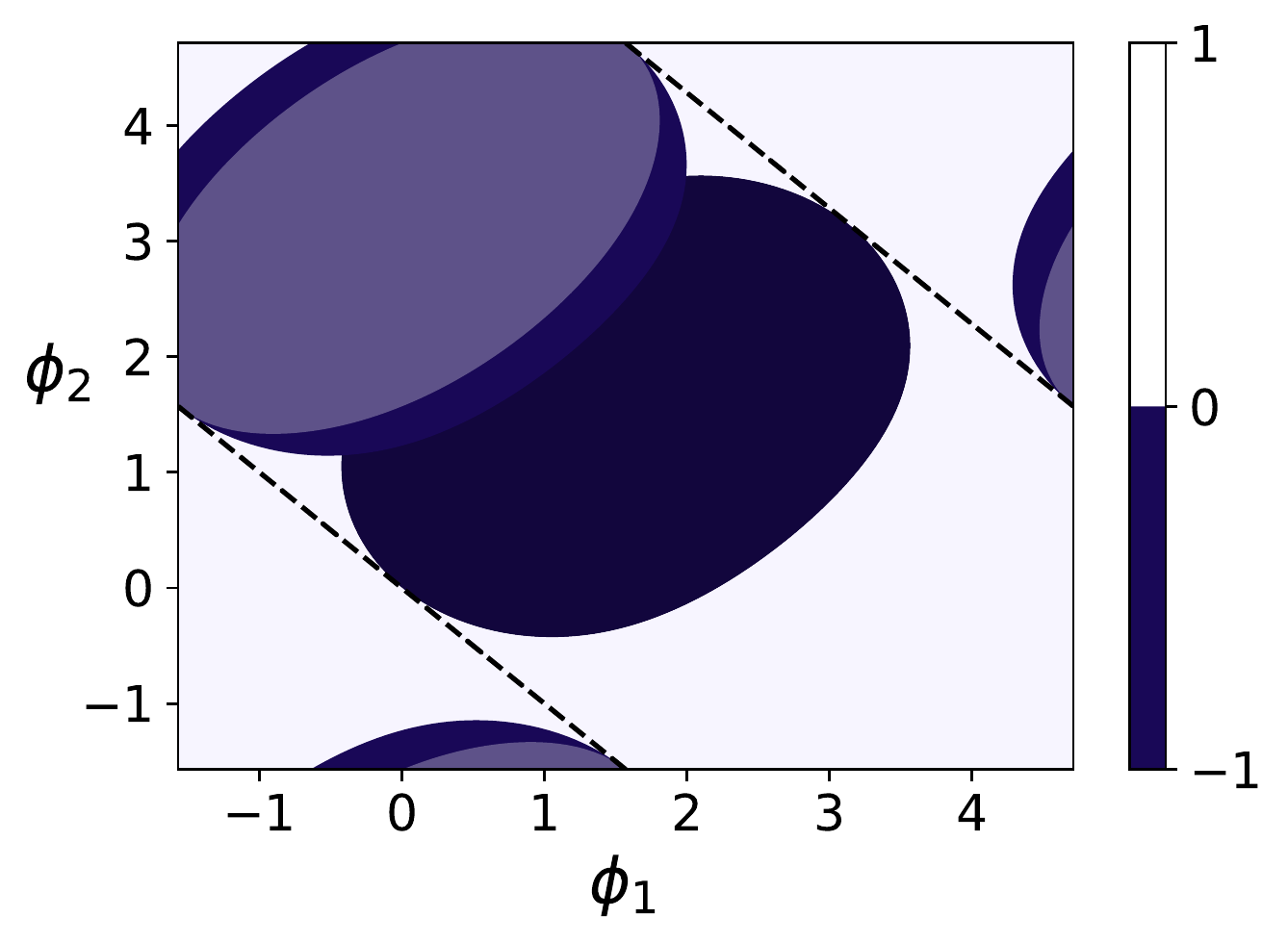}\label{fig:WAll3}}
    \caption{Regions of entanglement for (a) $W_{1}$, (b) $W_{2}$, and (c) $W_{3}$ for the state $\ket{\psi(\phi_{1},\phi_{2})}$ over a $2\pi \times 2\pi$ region in $\phi_{1}$,$\phi_{2}$ space.
    Purple regions show a negative expectation value, when the witness will register entanglement. Pale regions show a positive expectation value, where no entanglement can be deduced.
    Black dotted lines show when $\phi_{1}+\phi_{2} = 2n\pi$ where $n\in \mathbb{Z}$, i.e. when $\ket{\psi(\phi_{1},\phi_{2})}$ is in a product state.
    Each plot is suitably centred so that the size of the region in which entanglement can be detected is clear.
    Each pattern repeats periodically in both $\phi_{1}$ and $\phi_{2}$ with period $2\pi$.
    $W_{1}$ and $W_{2}$ do not detect entanglement for phase values close to $\phi_{1}=\phi_{2}=0$, but $W_{3}$ does. $W_{2}$ and $W_{3}$ have near identical detection areas as both can be considered as fidelity witnesses constructed from maximally entangled states.
    See Table~\ref{tbl:area} for a calculation of the proportion of the areas which detect entanglement for each entanglement witness. The entanglement regions for all three witnesses are shown simultaneously in (d) to show their relative sizes and displacements.}
    \label{fig:3Witnesses}
\end{figure*}

Instead of the above witness, we can construct a fidelity witness from a maximally entangled state. 
The eigenstate corresponding to the negative eigenvalue of \eqref{eq:W1p} is $\ket{\psi(0,\pi)}$, which is maximally entangled. 
If we construct a fidelity witness from this state we get
\begin{align}
    W_{2}&= \frac{1}{2}\mathbb{I} - \ketbra{\psi(0,\pi)}{\psi(0,\pi)} \nonumber \\
    &\propto I\otimes I + X\otimes Z - Z\otimes X + Y\otimes Y.
\end{align}
The expectation value $\expect{W_{2}}$ for the state \eqref{eq:psi} is shown in Fig.~\ref{fig:Wprime}. Since it is constructed from a maximally entangled state, we see it has a larger detection area than for $W_{1}^{\prime}$. 
We also see it requires an extra spin measurement as compared to $W_{1}^{\prime}$. However, it is still not optimised to detect small entanglement. A large amount of entanglement is required before $W_{2}$ will register it; assuming the experimental values in \cite{Bose2017}, this corresponds to a long interaction time.

In \cite{Chevalier2020} the authors provided an improved entanglement witness, given by
\begin{equation}
    W_{3} = I\otimes I - X\otimes X - Y\otimes Z - Z\otimes Y.\label{eq:WChev}
\end{equation}
The authors constructed this witness following a sequence of applications of the partial transpose operation on the state $\ket{\psi(\phi_{1},\phi_{2})}$, and then setting $\phi_{1}=\phi_{2}=0$.
However since this witness has one negative eigenvalue, and all positive eigenvalues are equal, it is proportional to witnesses of the form \eqref{eq:W}. The largest eigenvalue of this witness is $\beta = 2$, and so $\beta d - \tr(W_{3}) = 4$. It turns out that if we put $W_{3}$ into the form of \eqref{eq:rescaled} we find that
\begin{equation}
    \frac{1}{4}W_{3} = \frac{1}{2} \mathbb{I} - \ketbra{\psi(\pi/2,\pi/2)}{\psi(\pi/2,\pi/2)}.
\end{equation}
The state $\ket{\psi(\pi/2,\pi/2)}$ is maximally entangled, hence $W_{3}$ has an optimal detection area of $\frac{1}{2}$. Furthermore, this witness will detect small entanglement [see Fig.~\ref{fig:WChev}]. In fact if we calculate 
\begin{equation}
    |\braket{\psi(\pi/2,\pi/2)}{\psi(0,0)}|^{2} = \frac{1}{2},
\end{equation}
we see that the state $\ket{\psi(\pi/2,\pi/2)}$ is one of the closest maximally entangled states to the initial state $\ket{\psi(0,0)}$, and so it is an optimal entanglement witness by the criteria described in Sec.~\ref{sec:optimal}.
\begin{table}[b!]
\centering
$\begin{array}{|c|c|}
     & \text{Proportion of area} \\
     & \text{that detects Entanglement} \\
     \hline
     \expect{W_{1}}&22.6272\% \\
     \expect{W_{2}}&29.5392\%\\
     \expect{W_{3}}&29.5528\%
\end{array}$
\caption{Proportion of area that shows entanglement for each of the three witnesses described in Sec.~\ref{sec:gw}. The witnesses are integrated over a region of area $2\pi \times 2\pi$, similar to that in Fig.~\ref{fig:3Witnesses}. 
$W_{2}$ and $W_{3}$ can be considered fidelity witnesses constructed from maximally entangled states, so they have the largest detection areas; $W_{3}$ has slightly more.}\label{tbl:area}
\end{table}

In Fig.~\ref{fig:3Witnesses} we show the expectation value of these three witnesses for the state $\ket{\psi(\phi_{1},\phi_{2})}$, showing for which phases each witness will show entanglement.
One way to compare these three witnesses is to calculate the detection area in the space of phases $\phi_{1}$, $\phi_{2}$. 
In Table~\ref{tbl:area}, we show the proportion of the area of the regions in which each witness shows entanglement, in a $2\pi\times 2\pi$ region of $\phi_{1}$, $\phi_{2}$ space. We see that the area of $W_{1}$ is clearly less as expected, and the areas of $W_{2}$ and $W_{3}$ to be nearly identical. This is also to be expected as they can both be considered to be fidelity witnesses constructed out of maximally entangled states, but $W_{1}$ cannot.

\subsection{Optimal Gravitational Witness}

It remains a question of interest as to whether one can further improve the entanglement detection for this experimental setup.
Let us apply the techniques from Sec.~\ref{sec:optimal}. 
We wish to construct a fidelity witness from a maximally entangled state. Since we expect the entanglement in the state $\ket{\psi(\phi_{1},\phi_{2})}$ to be small, we seek the closest maximally entangled state to the initial state $\ket{\psi(0,0)}$. Since this state is a product state, the closest maximally entangled state is not unique. 
In fact, from \eqref{eq:closest}, we see that there is a one-parameter family of closest maximally entangled states, parametrised by $\theta$,
\begin{equation}
    \ket{\theta} = \frac{1}{\sqrt{2}}\left(\ket{\psi(0,0)}+e^{i\theta}\ket{\psi(\pi,\pi)}  \right). \label{eq:cloestqubits}
\end{equation}
We saw earlier that $W_{3}$ can be considered a fidelity witness constructed from the state $\ket{\psi(\pi/2,\pi/2)}$, which corresponds here to a choice $\theta = 3\pi/2$.
We can see from Fig.~\ref{fig:WChev} that $W_{3}$ is optimised to detect entanglement for small \emph{positive} phases. Indeed in \cite{Chevalier2020}, the authors approximated $\phi_{1}=0$, and assumed $\phi_{2} \geq 0$. However, we see from \eqref{eq:phases} that for non-zero interaction time $\tau$, we always have $\phi_{1}< 0$, and if $\ell < \Delta X$ then $\phi_{2}<0$. For $\tau\ll 1$, we calculate that 
\begin{equation}
    \expect{W_{3}} = -(\phi_{1}+\phi_{2}) + \mathcal{O}(\tau^{2}),
\end{equation}
and so $W_{3}$ will not detect entanglement for small interaction times if $\ell < \Delta X$. 

\begin{figure}[t]
    \centering
    \subfigure[]{\includegraphics[scale=0.6]{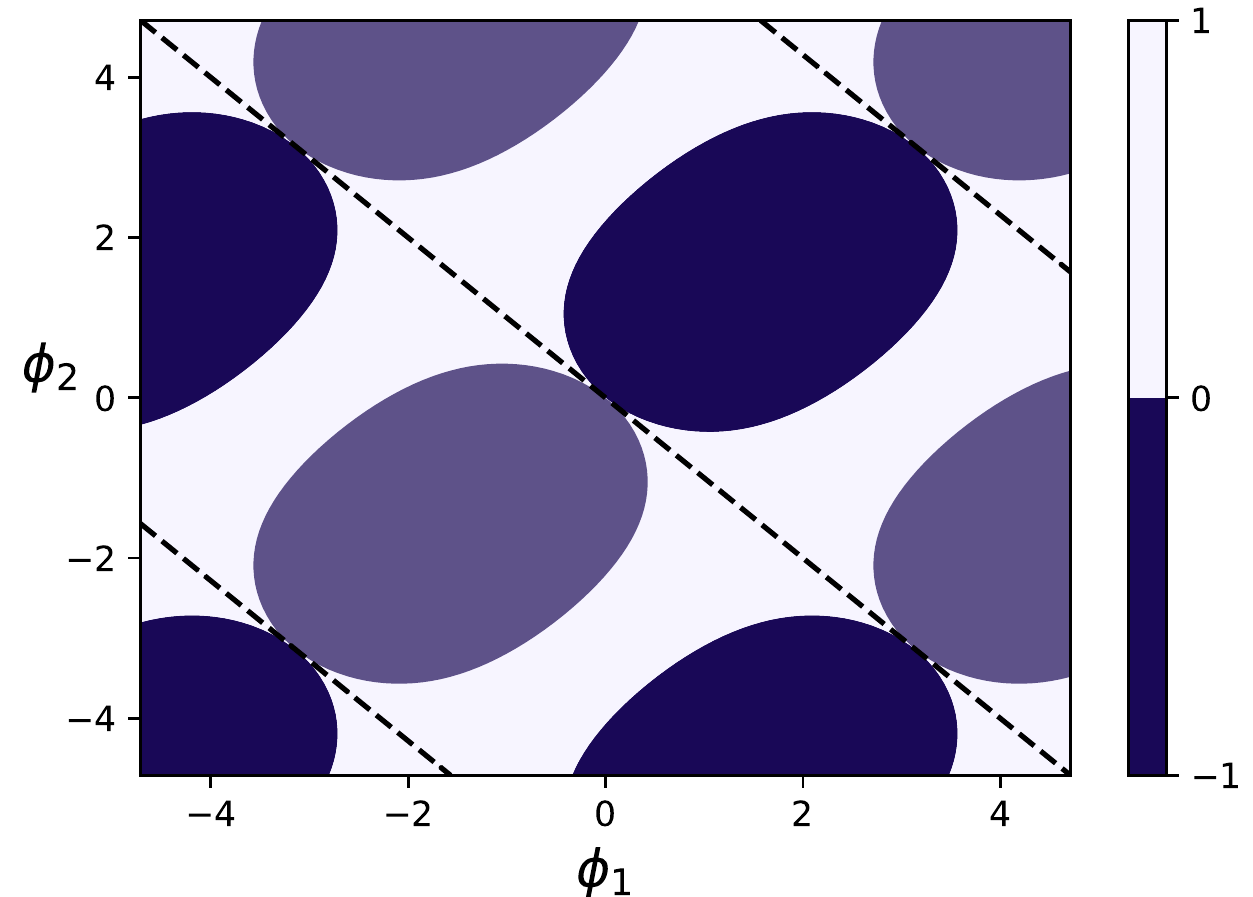}\label{fig:Chev2side}}
    \subfigure[]{\includegraphics[scale=0.6]{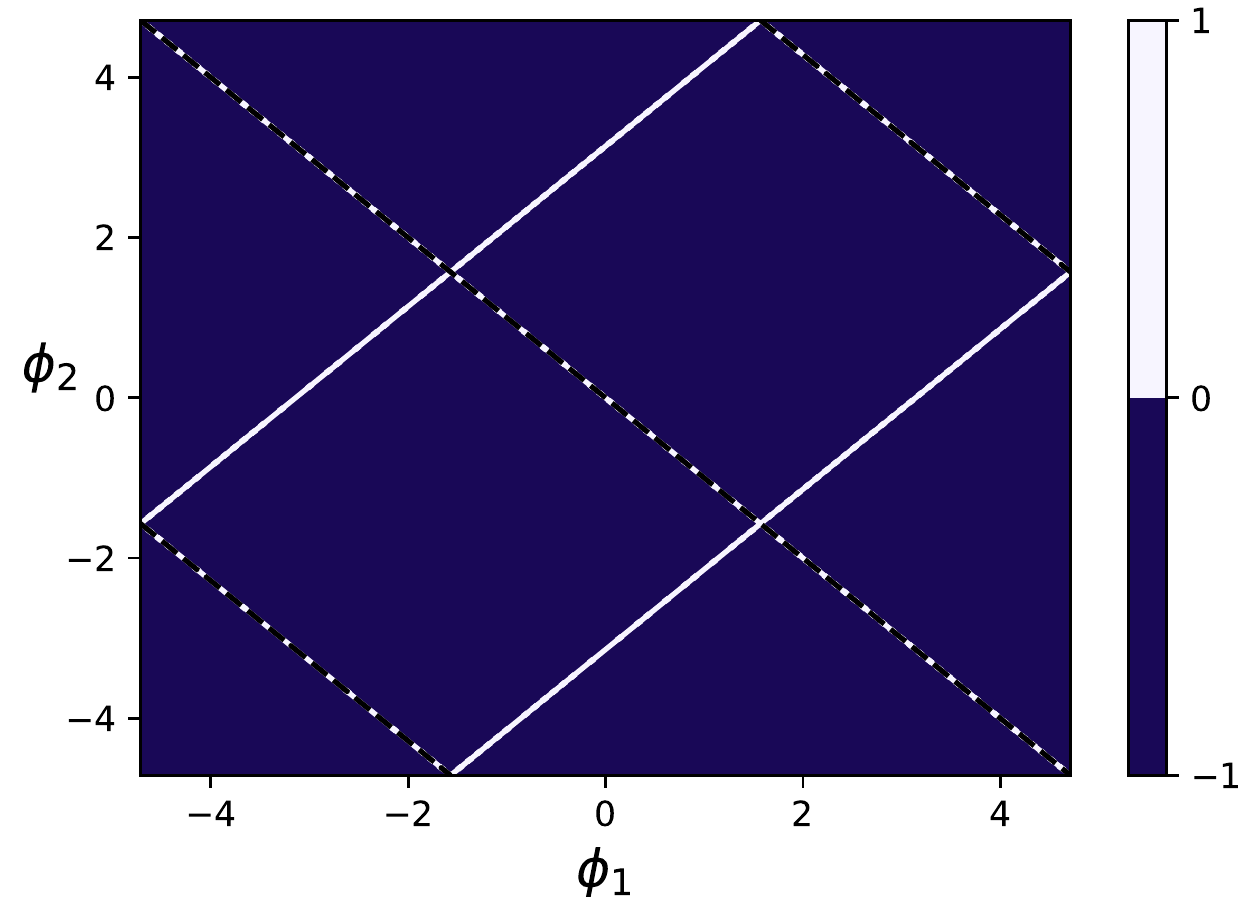}\label{fig:WGen}}
    \caption{The benefits of post-processing.\\
    (a) The combined region of detection for entanglement witnesses $W_{3}$ and $W_{4}$ for the state $\ket{\psi(\phi_{1},\phi_{2})}$. Dark purple regions indicate phases for which $\expect{W_{3}}$ is negative, lighter purple regions indicate phases for which $\expect{W_{4}}$ is negative, so the witness would detect entanglement. In the pale areas the expectation value of both is non-negative, so no conclusion about entanglement can be made using either witness. Black dotted lines show where $\phi_{1} + \phi_{2} = 2n\pi$, for $n\in \mathbb{Z}$, in which case the $\ket{\psi(\phi_{1},\phi_{2})}$ is in a product state. 
    Both $W_{3}$ and $W_{4}$ consist of the same local Pauli measurements, but with different post-processing, thus they can be measured with one set of measurements.
    (b) Expectation value of $W_{G}(\theta)$ for the same state, minimised over $\theta$ \eqref{eq:mintheta}. 
    This witness consists of five non-trivial local Pauli measurements, for any choice of $\theta$, and therefore post-processing can be used to apply the entanglement witness for any choice of $\theta$.
    Minimising over $\theta$ will detect entanglement for any choice of $\phi_{1}$ and $\phi_{2}$, except for when $\phi_{1}-\phi_{2} = (2n+1)\pi$ for $n\in \mathbb{Z}$: a subset set of measure zero.}
\end{figure}

If we instead construct a fidelity witness \eqref{eq:thetaGens} from the closest maximally entangled state with a value $\theta=\pi/2$, we find
\begin{align}
    W_{4} &= \frac{1}{2}\mathbb{I} - \ketbra{\pi/2}{\pi/2} \nonumber \\
    &\propto I\otimes I - X\otimes X + Y\otimes Z + Z\otimes Y.\label{eq:W4}
\end{align}
Comparing \eqref{eq:W4} to \eqref{eq:WChev} we see that $W_{4}$ uses the same spin measurements as $W_{3}$, but with different signs. However, the detection area of $W_{4}$ in $\phi_{1},\phi_{2}$ space is different from $W_{3}$, and indeed does not even overlap with it. Therefore with the same local measurements but different post-processing, one can double the detection area of the entanglement witness [see Fig.~\ref{fig:Chev2side}].
For small interaction times $\tau \ll 1$, we calculate
\begin{equation}
    \expect{W_{4}} \propto -\expect{W_{3}} + \mathcal{O}(\tau^{2}).
\end{equation}
Thus this witness will detect entanglement for the parameter regime when $\ell < \Delta X$. And therefore, in combination, $W_{3}$ and $W_{4}$ will be able to detect entanglement for small interaction times in both parameter regimes.
For this particular gravitational entanglement experiment, the regime $\ell < \Delta X$ may be unfeasible due to extra considerations such as the Casimir interaction. Nevertheless, this example shows how post-processing can greatly increase the parameter range for which a witness can detect entanglement.

Let us now construct the fidelity witness for the general state $\ket{\theta}$ \eqref{eq:cloestqubits}
\begin{align}
    W_{G}(\theta) &= \frac{1}{2}\mathbb{I} - \ketbra{\theta}{\theta} \nonumber \\
    &\propto  I\otimes I - X\otimes X + \cos\theta\,(Y\otimes Y - Z\otimes Z) \nonumber \\
    &\quad + \sin\theta\, (Y\otimes Z + Z\otimes Y).\label{eq:WG}
\end{align}
Thus $W_{G}(\theta)$ represents a one-parameter family of entanglement witnesses. 
The witnesses $W_{3}$ and $W_{4}$ can immediately be seen to be special cases of $\theta=3\pi/2$ and $\theta=\pi/2$, respectively.
The entire family of witnesses can be measured by measuring the five non-trivial spin operators in \eqref{eq:WG}, with post-processing to select the desired witness. Indeed, with the optimal post-processing (minimising $\expect{W_{G}(\theta)}$ over $\theta$), we can detect entanglement for the state $\ket{\psi(\phi_{1},\phi_{2})}$ for any values of $\phi_{1}$ and $\phi_{2}$, except for a set of measure zero. 

To see this, we calculate the expectation value of $W_{G}(\theta)$ in the state $\ket{\psi(\phi_{1},\phi_{2})}$,
\begin{align}
    \expect{W_{G}(\theta)}&\propto 2\sin^{2}\left(\frac{\phi_{1}-\phi_{2}}{2}\right)\sin^{2}\left(\frac{\theta}{2}\right) \nonumber \\
    &\qquad\quad+\sin\theta(\sin\phi_{1} + \sin\phi_{2}).\label{eq:WGexpect}
\end{align}
In order for this expectation value to be negative, we require
\begin{equation}
    2\sin^{2}\left(\frac{\phi_{1}-\phi_{2}}{2}\right) < \frac{-\sin\theta}{\sin^{2}\left(\frac{\theta}{2}\right)}(\sin\phi_{1} + \sin\phi_{2}).
\end{equation}
With the appropriate choice of $\theta$ in the interval $(0,2\pi)$, the ratio $-\sin\theta / \sin^{2}(\theta/2)$ can take on any real value. Hence one can choose $\theta$ such that the expectation value \eqref{eq:WGexpect} is always negative, unless $\sin\phi_{1} + \sin\phi_{2} = 0$. This will happen under two conditions:
\begin{equation}
    \phi_{1} + \phi_{2} = 2 n\pi \quad \text{or} \quad \phi_{1}-\phi_{2} = (2n+1)\pi,
\end{equation}
where $n\in \mathbb{Z}$. As we saw earlier, the first case corresponds to when the state $\ket{\psi(\phi_{1},\phi_{2})}$ is not entangled. 
In the second, the state may be entangled, but the expectation value with $W_{G}(\theta)$ \eqref{eq:WGexpect} will be zero for any choice of $\theta$. 
This means that the witness $W_{G}$ cannot detect entanglement in the state $\ket{\psi(\phi_{1},\phi_{2})}$ if $\phi_{1}-\phi_{2} = (2n+1)\pi$ for any choice of $\theta$, but the subset of phases satisfying this condition is a set of measure zero.

Therefore, by performing only five non-trivial spin measurements, and minimising over $\theta$,
\begin{equation}
    \min_{\theta\in [0,2\pi]} \expect{W_{G}(\theta)},\label{eq:mintheta}
\end{equation}
one can detect the entanglement for any choice of phases $\phi_{1}$ and $\phi_{2}$, except for the aforementioned set of measure zero (see Fig.~\ref{fig:WGen}). We note that the exact phases $\phi_{1}$ and $\phi_{2}$ will not be known due to experimental imperfections. Thus, this is a desirable measurement strategy, as it allows one to explore all possible entanglement signatures by post-processing the acquired measurement data.

\section{Dephasing}\label{sec:dephasing}

\begin{figure*}[t!]
    \centering
    \subfigure[]{\includegraphics[scale=0.43]{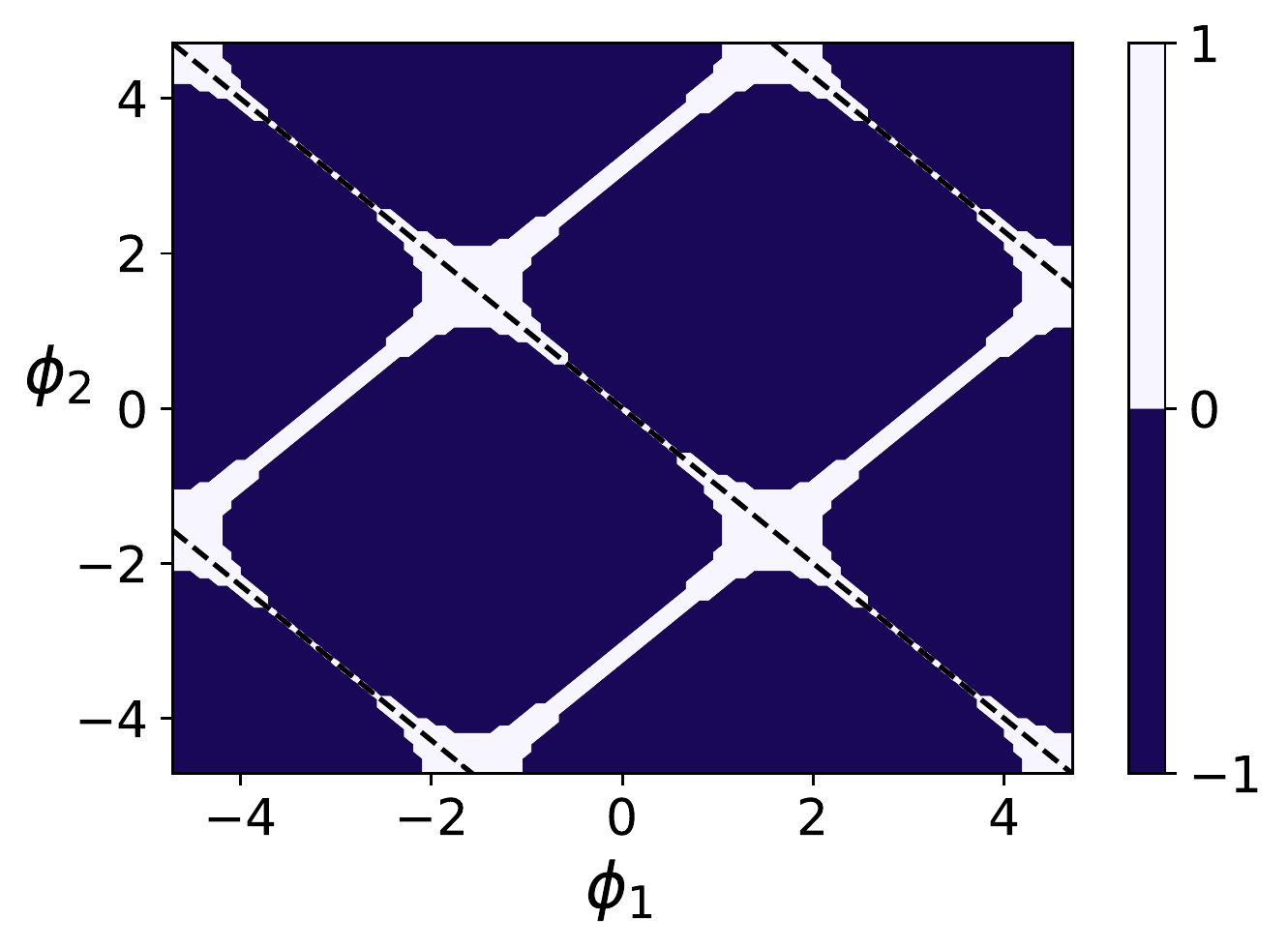}\label{fig:0.005}}
    \subfigure[]{\includegraphics[scale=0.43]{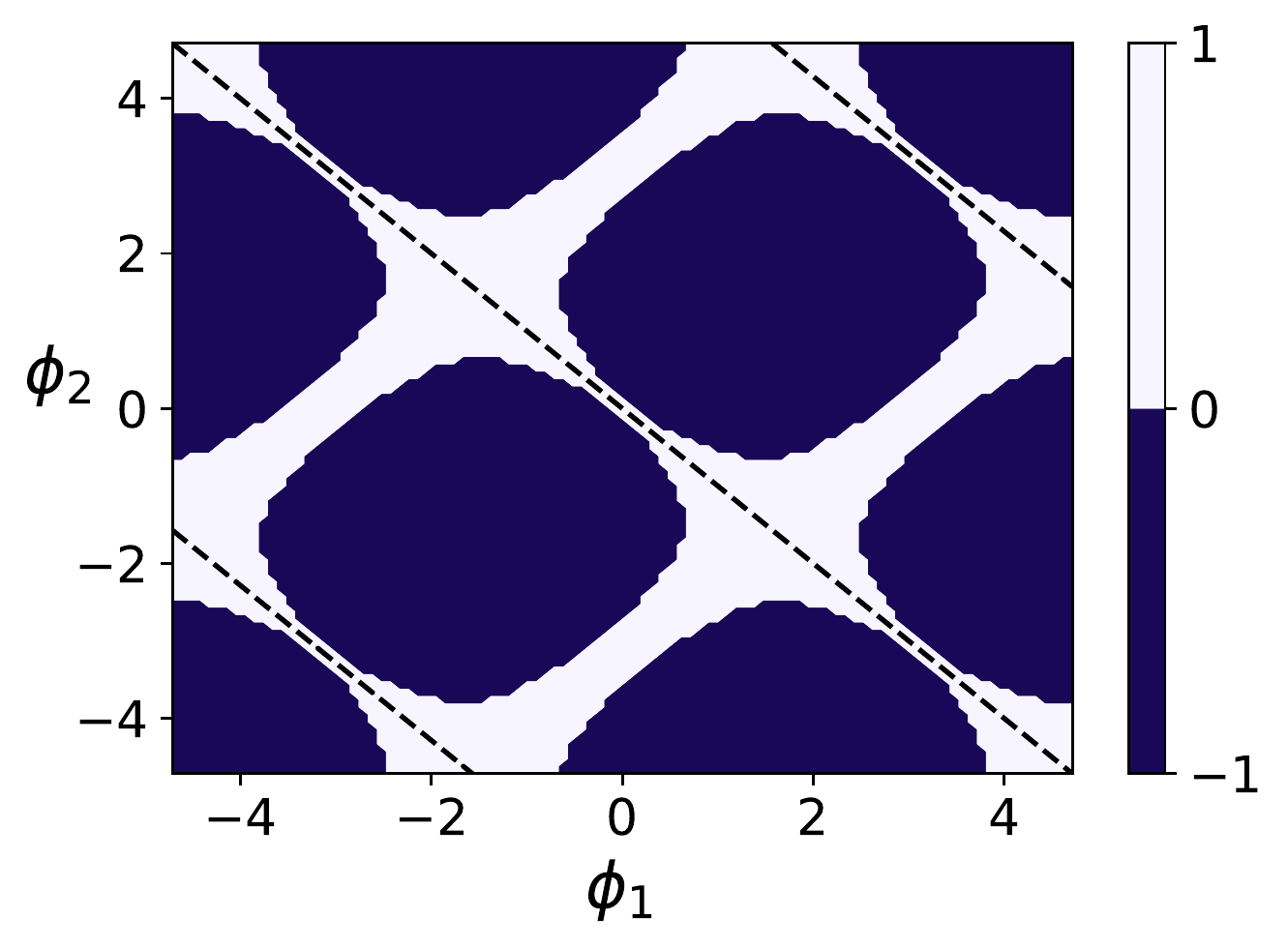}\label{fig:0.05}}
    \subfigure[]{\includegraphics[scale=0.43]{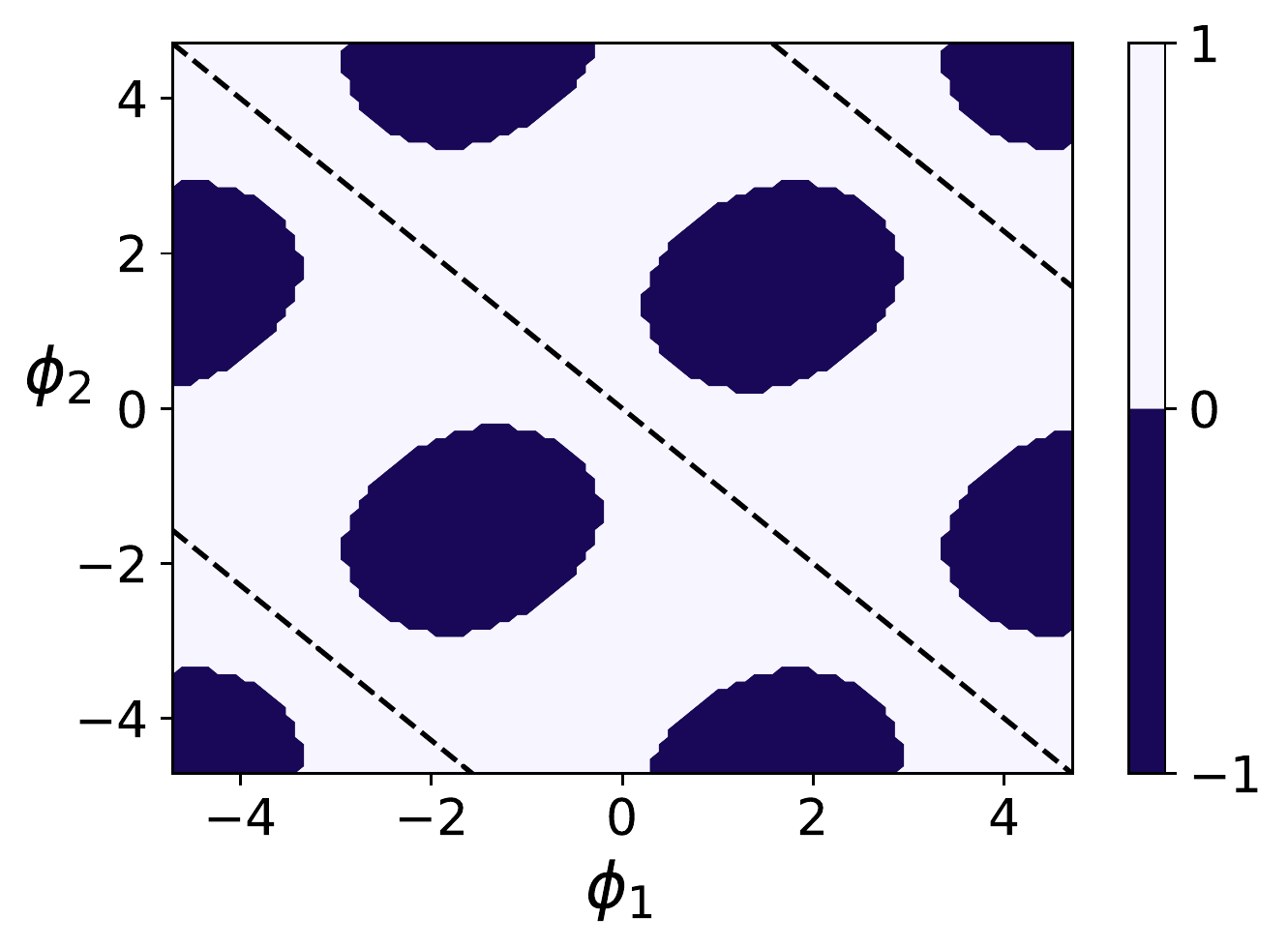}\label{fig:0.5}}
    \caption{Effects of dephasing \eqref{eq:dephchannel} on the expectation value of $W_{G}$. 
    Dark regions show a negative expectation value, when the witness will register entanglement. Light areas show a positive expectation value, where no entanglement can be deduced.
    Black dotted lines show when $\phi_{1}+\phi_{2} = 2n\pi$ where $n\in \mathbb{Z}$, i.e., when $\ket{\psi(\phi_{1},\phi_{2})}$ is in a product state.
    The expecation value is explicitly given by \eqref{eq:dephased}, minimised over $\theta\in [0,2\pi]$. Here we have plots for (a) $\gamma=0.005$, (b) $\gamma=0.05$ and (c) $\gamma=0.5$. As $\gamma$ increases the area in which the witness will recognise entanglement is reduced. When $\gamma=0$, $W_{G}$ can be used to detect entanglement for any choice of $\phi_{1},\phi_{2}$, except for a set of measure zero [see Fig.~\ref{fig:WGen}].} 
    \label{fig:dephasing}
\end{figure*}

It is important to see the effects of noise on the system, since all real experiments will have some loss associated with interaction with an environment. 
There is recent work trying to use larger dimensional systems and different geometries to reduce the effects of decoherence \cite{tilly2021qudits}.

As in \cite{Chevalier2020}, with no explicit model for the environment, we assume a simple dephasing channel on each system,
\begin{equation}
    \mathcal{E}(\rho) = (1-p)\rho + p Z\rho Z,\label{eq:dephchannel}
\end{equation}
where $p=\frac{1}{2}(1-e^{-\gamma})$. In the limit $\gamma\rightarrow \infty$ this channel causes full dephasing.
If we now calculate the expectation value of $W_{G}(\theta)$ with the dephased state $\mathcal{E}(\ketbra{\psi(\phi_{1},\phi_{2})}{\psi(\phi_{1},\phi_{2})})$, we find
\begin{align}
    \expect{W_{G}(\theta)}_{\gamma} &\propto 1+e^{-\gamma}\sin\theta (\sin \phi_{1} + \sin\phi_{2})\label{eq:dephased} \\ &-e^{-2\gamma}\left(1-2\sin^{2}\left(\frac{\theta}{2}\right)\sin^{2}\left(\frac{\phi_{1}-\phi_{2}}{2}\right)\right)\nonumber.
\end{align}
If we compare this with \eqref{eq:WGexpect} we see an exponential suppression of the region of detectable entanglement. In Fig.~\ref{fig:dephasing} we  plotted the effect of the local dephasing channel on the witness $W_{G}(\theta)$ with the same post-processing as in \eqref{eq:mintheta}, for three values of $\gamma$ at different orders of magnitude; showing that the entire family of witnesses become less effective at detecting entanglement with larger dephasing.

\section{Generality}\label{sec:disc}

In Sec.~\ref{sec:ew} we restricted consideration to entanglement witnesses which detect entanglement only with a negative expectation value. This was mathematically useful since rescaling such a witness by a positive constant did not change the set of states for which the witness would detect entanglement.
Other witnesses can often be brought into this form by reflection (rescaling by $-1$), and/or translation [adding multiples of the identity, such as \eqref{eq:W1p}]. However we saw an example of a witness $W_{1}$ which could not be brought into such a form, as its set of detection values consisted of two disjoint open regions of the real line.
To analyse it in our framework, we only focused on the half of the set of detection values which was relevant for the experimental proposal to detect gravitational entanglement \cite{Bose2017}.
Thus we studied the related witness $W_{1}^{\prime}$ \eqref{eq:W1p}, and assigned it a suboptimal detection area of $\frac{1}{8}$.
To study the other half of the witness, we can reflect and translate $W_{1}$ to
\begin{equation}
    W_{1}^{\prime\prime} = I \otimes I - X\otimes Z - Y\otimes Y. \label{eq:W1pp}
\end{equation}
The original witness $W_{1}$ can be written as $\frac{1}{2}(W_{1}^{
\prime}-W_{1}^{\prime\prime})$.
Unsurprisingly this witness $W_{1}^{\prime\prime}$ also has a largest eigenvalue of $\beta=3$, and the corresponding density operator $\rho$ has a largest eigenvalue of $\frac{1}{2}$, so the detection area of this witness is also $\frac{1}{8}$. So in total, the witness $W_{1}$ should be assigned a detection area of $\frac{1}{4}$. Thus we see that a way to measure the detection area of an entanglement witness with disjoint sets of detection values is to split the witness up into multiple witnesses and sum their respective detection areas.

In this article we applied this framework to the experimental proposal described in \cite{Bose2017}, which involved detecting entanglement of two qubits. This is the simplest such application as the closest maximally entangled state to a product state could be described by a one-parameter family of states. For higher dimensions this will no longer hold and describing the closest maximally entangled state will be more involved. 

The next step to continue this line of research is to generalise this framework. First, fidelity witnesses constructed for mixed entangled states could be better understood. While we saw that the witness $\eqref{eq:W}$ was an entanglement witness whenever $\ket{\psi}$ was entangled, in general the operator $\alpha \mathbb{I} - \rho$ need not be an entanglement witness even when $\rho$ is entangled. It will be certainly true (with a suitable choice of $\alpha$) in the specialised case when the largest eigenvalue of $\rho$ is non-degenerate and the corresponding eigenspace is spanned by an entangled state.
We argued that the best fidelity witness should be one constructed from one of the closest maximally entangled states from the state of interest. In this article we constructed this state for pure states using the Schmidt decomposition. It would be useful to calculate this for entangled mixed states.

Another generalisation is to multi-partite entanglement. 
Fidelity witnesses have the advantage over other witness constructions of being easy to generalise to multi-partite entanglement \cite{Guhne2009}.
However bipartite entanglement allows the use of the helpful Schmidt decomposition, which allows one to calculate the coefficient $\alpha$ \eqref{eq:alpha} in terms of Schmidt coefficients; and allowed us to characterise the closest maximally entangled state. Since the Schmidt decomposition does not, in general, hold for higher multi-partite entanglement, performing the above calculations will be more challenging.

\section{Conclusion}\label{sec:conc}

Confirming that a quantum state is entangled is in general a challenging task. Entanglement witnesses make this task easier by reducing the task to calculating or measuring the expectation value of an observable.
Unfortunately, no entanglement witness can detect entanglement for all entangled states. Thus if an entanglement witness does not tell us that a state is entangled, we cannot conclude that a state is in fact not entangled.
This means that for a particular experimental setup, one needs to find an entanglement witness best suited to detect entanglement.

In this article we looked at the recent experimental proposal to detect gravitational induced entanglement \cite{Bose2017} using fidelity witnesses.
Fidelity witnesses offer a natural measure of the volume of states for which they detect entanglement.
We showed that this measure can be generalised to a much broader class of witnesses by rewriting them into an analogous form \eqref{eq:rescaled}.
By this measure, entanglement witnesses constructed from pure, maximally entangled states will detect entanglement from the greatest volume of entangled states.
Therefore we argue that an optimal choice for an entanglement witness is a fidelity witness constructed from one of the closest maximally entangled states to the state of interest, measured by the fidelity.
We found a general formula for such a state when the state of interest is a bipartite pure state, which we applied to the experimental proposal in \cite{Bose2017}. 
In that case the closest maximally entangled state to the initial product state was a one-parameter family of states, and minimising over this parameter leads to the effective entanglement witness displayed in Fig.~\ref{fig:WGen}.
In other words, with the appropriate post-processing, this witness can detect entanglement for almost any phases $\phi_{1},\phi_{2}$. We saw that the ability of these witnesses to detect entanglement decreases with larger dephasing.

This witness can be compared to others proposed for the same experiment. The original witness proposed in \cite{Bose2017} required a large amount of entanglement in the state before it is detected by the witness. A better witness that was suggested in \cite{Chevalier2020} detected smaller entanglement, and can be seen to be a specific example of a fidelity witness constructed from one of the closest maximally entangled states. As such it only detected a particular region of the configuration space.

We expect our presented method of constructing witnesses to be useful for other experiments which measure entanglement; whereby this construction produces witnesses better tailored to those particular experimental setups.

\section{Acknowledgements}
This work was supported by the Swedish Research
Council under grant no. 2019-05615, the European Research Council under grant no. 742104 and The Branco Weiss Fellowship – Society in Science.

\bibliography{references}

\end{document}